\documentclass[prl,superscriptaddress,showpacs,preprintnumbers,amsmath,amssymb,twocolumn,floats]{revtex4}
\usepackage{txfonts}
\usepackage{amssymb}
\usepackage{graphicx}
\usepackage[]{sidecap}
\usepackage{CJK}
\usepackage{txfonts}
\usepackage{pifont}

\begin{document}

\title{Signature of strong spin-orbital coupling in the large non-saturating magnetoresistance material WTe$_2$}

\author{J. Jiang }

\affiliation{State Key Laboratory of Surface Physics, Department of Physics, and Advanced Materials Laboratory, Fudan University, Shanghai 200433, China}

\affiliation{Collaborative Innovation Center of Advanced Microstructures, Fudan University, Shanghai 200433, China}

\author{F. Tang}

\author{X. C. Pan }

\author{H. M. Liu}

\affiliation{Department of Physics, Nanjing University,  Nanjing 210093, China}
\affiliation{Collaborative Innovation Center of Advanced Microstructures, Nanjing University, Nanjing 210093, China}

\author{X. H. Niu }

\author{Y. X. Wang}

\author{D. F. Xu}

\affiliation{State Key Laboratory of Surface Physics, Department of Physics, and Advanced Materials Laboratory, Fudan University, Shanghai 200433, China}

\affiliation{Collaborative Innovation Center of Advanced Microstructures, Fudan University, Shanghai 200433, China}

\author{H. F. Yang }

\affiliation{State Key Laboratory of Functional Materials for Informatics, Shanghai Institute of Microsystem and Information Technology (SIMIT), Chinese Academy of Sciences, Shanghai 200050, China}

\author{B. P. Xie }

\affiliation{State Key Laboratory of Surface Physics, Department of Physics, and Advanced Materials Laboratory, Fudan University, Shanghai 200433, China}

\affiliation{Collaborative Innovation Center of Advanced Microstructures, Fudan University, Shanghai 200433, China}

\author{F. Q. Song }

\author{X. G. Wan}

\affiliation{Department of Physics, Nanjing University, Nanjing 210093, China}
\affiliation{Collaborative Innovation Center of Advanced Microstructures, Nanjing University, Nanjing 210093, China}

\author{D. L. Feng }

\email{dlfeng@fudan.edu.cn}

\affiliation{State Key Laboratory of Surface Physics, Department of Physics, and Advanced Materials Laboratory, Fudan University, Shanghai 200433, China}

\affiliation{Collaborative Innovation Center of Advanced Microstructures, Fudan University, Shanghai 200433, China}

\begin{abstract}

We report the detailed electronic structure of WTe$_2$ by high resolution angle-resolved photoemission spectroscopy. Unlike the simple one electron plus one hole pocket type of Fermi surface topology reported before, we resolved a rather complicated Fermi surface of WTe$_2$. Specifically, there are totally nine Fermi pockets, including one hole pocket at the Brillouin zone center $\Gamma$, and two hole pockets and two electron pockets on each side of $\Gamma$ along the $\Gamma$-$X$ direction. Remarkably,  we have observed  circular dichroism  in our photoemission spectra, which suggests that  the  orbital angular momentum exhibits a rich  texture at various sections of the Fermi surface. As reported previously for topological insulators and Rashiba systems, such a circular dichroism is a signature for spin-orbital coupling (SOC). This is further confirmed by our density functional theory calculations, where the spin texture is qualitatively reproduced as the conjugate consequence of SOC. Since the backscattering processes are directly involved with the resistivity, our data suggest that the SOC and the related spin and orbital angular momentum textures may be considered in the understanding of the anomalous magnetoresistance of WTe$_2$.

\end{abstract}

\maketitle


\begin{figure}
\includegraphics[width=8.7cm]{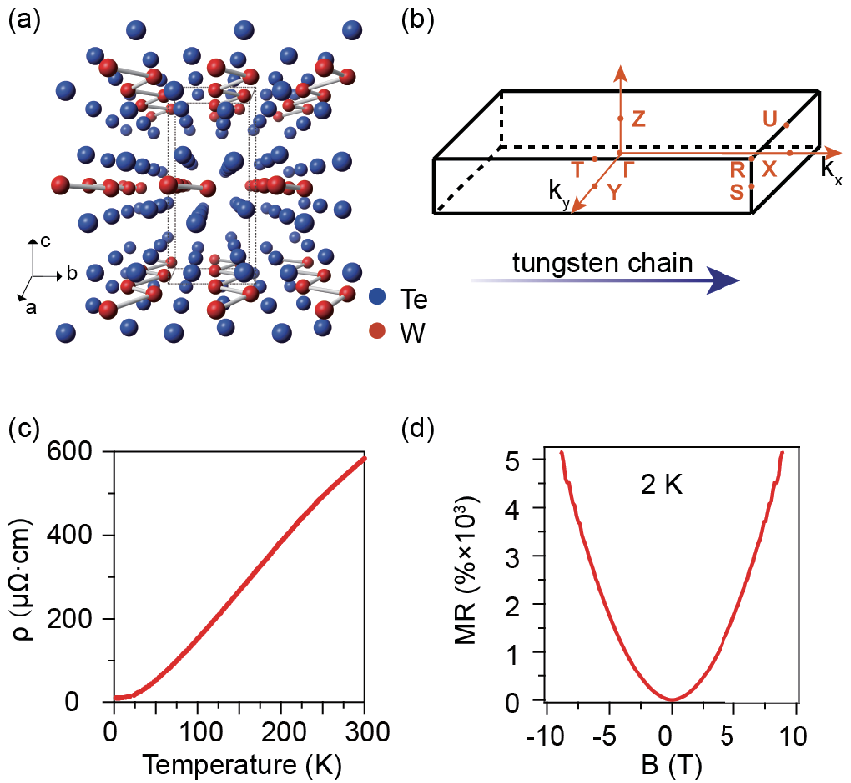}
\caption{(Color online) (a) Crystal structure of WTe$_2$, showing the layered structure. The blue spheres refer to the Te atoms and the red ones refer to the W atoms. (b) Three dimensional Brillouin zone of WTe$_2$. The high symmetry points are indicated. (c) Temperature dependent resistivity of WTe$_2$ single crystals measured from 2~K to 300~K. (d) The MR of WTe$_2$ single crystal up to 9~T, measured at 2~K.
} \label{structure}
\end{figure}

The magnetoresistance (MR) of a material reflects the dynamics of the charge carriers and the characteristics of the Fermi surface, and it can be useful for  magnetic memory and spintronics devices\cite{Pippard, Wolf}. Normally, one might expect MR varies as B$^2$ under low fields, and saturates under high fields. However, anomalously large MR has been discovered in some non-magnetic materials, such as Ag$_2$Te/Se\cite{AgTe1, AgTe2}, Bi\cite{Bi1, Bi2}, Bi$_2$Te$_3$\cite{BiTe}, Cd$_3$As$_2$\cite{Cava_CdAs}, and recently reported in WTe$_2$ and NdSb$_2$ \cite{Cava_WTe,NdSb}. MR is often very complicated, involving various factors, and  there are several possible mechanisms for these exotic MR behaviors. For example, the large linear MR in Bi$_2$Te$_3$ and Bi$_2$Se$_3$ is attributed to the quantum limit of the Dirac fermions, namely all the carriers are situated in the first Landau level \cite{Abrikosov}. For both Bi and PtSn$_4$,  the large MR is attributed  to the equal amount of electrons and holes, in other words, due to Fermi surface compensation \cite{Bi1, PtSn}. Intriguingly, the large MR in the Dirac semi-metal Cd$_3$As$_2$ is attributed to certain protection mechanism that strongly suppresses backscattering in zero field; such a mechanism would be progressively invalidated by the presence of magnetic field,  thus causing large MR \cite{Cava_CdAs}.

WTe$_2$ is recently reported to exhibit extremely large MR, which shows no signature of saturation up to 60~T \cite{Cava_WTe}. It was suggested that such an exotic behavior can be understood in the framework of Fermi surface compensation \cite{Cava_WTe}. Consistently, angle-resolved photoemission spectroscopy (ARPES) did observe one electron and one hole Fermi pocket with equal area on each side of  the Brillouin zone (BZ) center $\Gamma$ \cite{Valla}. Further evidence comes from the drastically reduced MR with increased pressure up to 23.6~kPa, which is accompanied by the increasing size difference between the electron and hole pockets as observed in the Subnikov de Hass (SdH) oscillation experiments \cite{SYLi}. Interestingly, superconductivity emerges when the pressure is further enhanced with a maximal transition temperature of about 7~K \cite{SC, LLSun}.

\begin{figure*}[t]
\includegraphics[width=12.5cm]{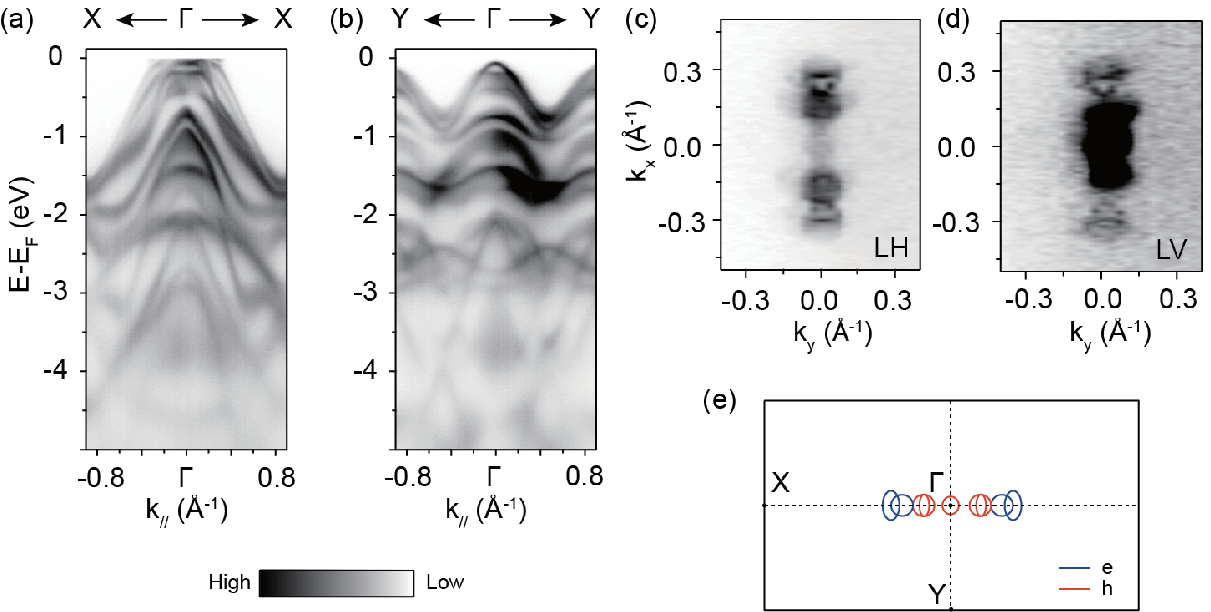}
\caption{(Color online) (a,b), Large scale band structures of WTe$_2$ along the $\Gamma$-$X$ and $\Gamma$-$Y$ directions, respectively. (c,d), Fermi surface mapping under linear-horizontal (LH) and linear vertical (LV) polarized light, respectively. (e) Schematic of the Fermi surface of WTe$_2$, where the red lines indicate the hole pockets and the blue lines indicate the electron pockets. Data were taken at 7~K with 58~eV photons.
} \label{FS}
\end{figure*}

\begin{figure*}
\includegraphics[width=18cm]{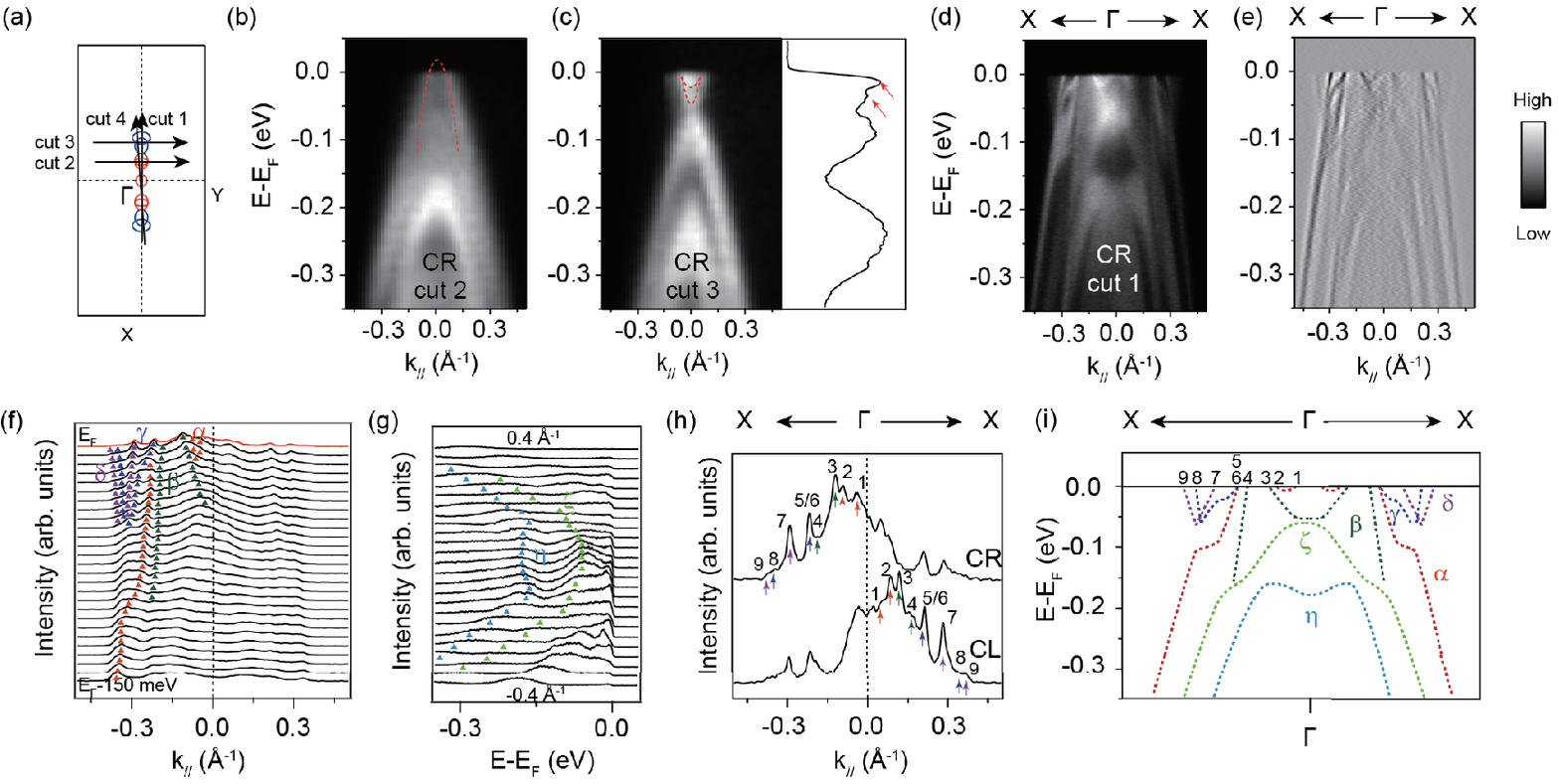}
\caption{(Color online) (a) Schematic of the Fermi surface of WTe$_2$ and the different cut directions are indicated by the black arrows. (b) The photoemission intensity plot along cut 2. The red dashed line indicates the dispersion of the hole band. (c) Left: The photoemission intensity plot along cut 3. The red dashed lines indicate the dispersion of the electron bands. Right: The corresponding EDC where the two red arrows highlight the band bottom of the two electron bands. (d,e), The photoemission intensity plot along $\Gamma$-$X$ together with its corresponding second derivative spectrum. (f) The corresponding MDC's of the spectrum in panel (d) in an energy window [$E_F$-150 meV, $E_F$]. (g) The corresponding EDC's of the spectrum in panel (d) in a momentum window [-0.4 $\AA^{-1}$, 0.4 $\AA^{-1}$]. (h) MDC's at $E_F$ under RCP light and LCP light, respectively. (i) Schematic of the low-lying electronic structure of WTe$_2$. Data were taken at 7~K with 58~eV photons.
}\label{cut}
\end{figure*}

Considering the complexity of MR, we examine other factors that may contribute to the anomalous MR behavior of WTe$_2$ in this Letter. We have conducted  ARPES measurements on WTe$_2$ single crystals and observed more subtle yet critical details of the electronic structure. More specifically, we have observed nine Fermi pockets instead of four reported earlier  \cite{Valla}, and we found that the photoemission intensity exhibits a strong dependence on the polarization of the incident photons. There is a remarkable intensity inversion between the data taken with right-circular polarized (RCP) and left-circular polarized (LCP) light. Such a circular dichroism (CD) suggests that the orbital angular momentum (OAM) varies with Fermi momentum, an evidence for the presence of strong spin-orbital coupling (SOC). This is further confirmed qualitatively by our band calculations, where the spin texture is observed, like in the topological insulators \cite{JJ_SmB6,CD1}.  The observed opposite OAM and spin direction on the opposite side of $\Gamma$ would provide a mechanism to protect the backscattering in zero field. When a magnetic field is imposed, such a mechanism would be compromised progressively, and thus affects the resistivity. Therefore, as suggested in the case of Cd$_3$As$_2$ \cite{Cava_CdAs},  such a SOC induced effects may play a role in the large non-saturating MR of WTe$_2$.

High quality WTe$_2$ single crystals were synthesized by a chemical vapor transport method as reported elsewhere\cite{SC}. The in-house ARPES measurements were performed with SPECS UVLS discharge lamp (21.2 eV He-I$\alpha$ light), and a Scienta R4000 electron analyzer. The synchrotron ARPES measurements were performed at the I05 beamline of Diamond Light Source (DLS) equipped with a Scienta R4000 electron analyzer and at the APE beamline of Elettra Sincrotrne Trieste equipped with a Scienta DA30 electron analyzer. The angular resolution was 0.3$^\circ$ and the overall energy resolution was better than 15~meV depending on different photon energies and beamlines. The samples were cleaved $in situ$ along the (001) plane and measured under ultra-high vacuum below 5$\times$10$^{-10}$ torr.

The electronic band structure calculations have been carried out by using the WIEN2K code with a full potential linearized augmented plane wave method and together with the Perdew-Burke-Ernzerhof parameterization of the generalized gradient approximation as the exchange-correlation functional \cite{cal1, cal2}. The basic functions are expanded to R$_{mt}$K$_{max}$ = 7 (where Rmt is the smallest of the muffin-tin sphere radii and Kmax is the largest reciprocal lattice vector used in the plane-wave expansion and the Brillouin zone was sampled by 10,000 k points. Using the second-order variational procedure, we include the SOC interaction.

WTe$_2$ shares a layered structure as other transition-metal dichalcogenides, but with an additional structural distortion. As shown in Fig.~\ref{structure}(a), the distorted metal layers are sandwiched between adjacent distorted dichalcogenide layers, and stacked along the $c$ axis through van der Waals bonding. Thus tungsten chains form the $a$ axis within the dichalcogenide layers. The BZ of WTe$_2$ is plotted in Fig.~\ref{structure}(b), where the tungsten chain direction defines the $\Gamma$-$X$ direction, and the perpendicular direction gives $\Gamma$-$Y$. The temperature dependent resistivity from 2~K to 300~K is plotted in Fig.~\ref{structure}(c), giving the relative residual resistivity (RRR) around 58. The MR of our sample shows no signature of saturation with the value of 5000$\%$ at 2~K under 9~T. More data of the large non-saturating behavior at higher magnetic filed of our samples can be found elsewhere \cite{XCPan}.

The large scale electronic structure of WTe$_2$ along $\Gamma$-$X$ and $\Gamma$-$Y$  are shown in Figs.~\ref{FS}(a) and (b), respectively. These bands are mainly contributed by the W $5d$ orbital and Te $5p$ orbital \cite{Augustin}. The band structure shows strong anisotropy along $\Gamma$-$X$ and $\Gamma$-$Y$  which are more obvious in the polarization dependent data of the Fermi surface  mapping presented in Figs.~\ref{FS}(c) and (d). There is a hole pocket around the BZ center $\Gamma$, whose intensity is more enhanced in the linear-vertical (LV) polarization. Furthermore, a rather complicated Fermi surface topology is observed along the tungsten chain direction ($\Gamma$-$X$). As will be elaborated later, five hole pockets and four electron pockets are resolved  along the $\Gamma$-$X$ direction. The Fermi surface observed are summarized in the Figs.~\ref{FS}(e). Since the charge carriers can be scattered easily along the $\Gamma$-$X$ direction, this might be responsible for the anisotropic resistivity.


The photoemission intensity plots along cut 2 and cut 3 are shown in Figs.~\ref{cut}(b) and (c),  respectively. The two electron bands are degenerate at Fermi energy ($E_F$) across cut 3, but we can distinguish them from the double peaks of their band bottoms in the corresponding energy distribution curve (EDC) in Fig.~\ref{cut}(c). The band structure along $\Gamma$-$X$ is rather complicated as shown in Fig.~\ref{cut}(d), together with its second derivative spectrum shown in Fig.~\ref{cut}(e). The corresponding momentum distribution curves (MDC's) and EDC's are shown in Figs.~\ref{cut}(f) and (g), respectively. There are totally four bands across the Fermi level, referred as $\alpha$, $\beta$, $\gamma$ and $\delta$. The $\alpha$ band totally contributes three Fermi crossings along $\Gamma$-$X$, crossing 1, 2 and 5 as indicated in Fig.~\ref{cut}(h), which forms the two hole pockets. The $\beta$ band gives two Fermi crossings indicated as crossing 3 and crossing 4, which forms the other hole pocket. The two electron bands, $\delta$ and $\gamma$, cross below $E_F$ with Fermi crossings marked as crossing 6 to crossing 9, which form the two electron FSs shown in Fig.~\ref{FS}(e). There are two bands located around 60~meV below $E_F$ (the $\eta$ band) and around 185~meV below $E_F$ (the $\zeta$ band).

The low-lying electronic structure observed along the $\Gamma$-$X$ direction are summarized in Fig.~\ref{cut}(i), which reveals  more details than the earlier ARPES work \cite{Valla}, where only two electrons and two hole pockets are resolved along $\Gamma$-$X$. To estimate the ratio of the electron and hole pockets, the $k_z$ dependence of the bands need to be taken into account. However, all the bands observed here didn't show obvious photon energy dependence [see supplementary material (SM)]. Despite the complexity of the resolved Fermi surface, we roughly estimate the ratio of the hole and electron pockets which is around 89$\%$, regardless of the $k_z$ dependence of the bands.



It has been shown that the OAM of an electronic state would manifest itself in the CD of ARPES spectra. Particularly, when there strong SOC is present, although total angular momentum is conserved, both OAM and spin angular momentum of a state would exhibit conjugate textures around the Fermi surface. For example, the topological surface state in Bi$_2$Se$_3$ and SmB$_6$ exhibit obvious chirality of OAM in their ARPES data \cite{CD1, JJ_SmB6}. In spin-orbital coupled systems, the photoemission CD signal is proportional to the inner product between OAM and light propagation vector \cite{CD1}. The CD-ARPES technique has been demonstrated to be a powerful tool to investigate the OAM texture of the surface states in TIs\cite{CD1,CD2}.

\begin{figure}
\includegraphics[width=8.7cm]{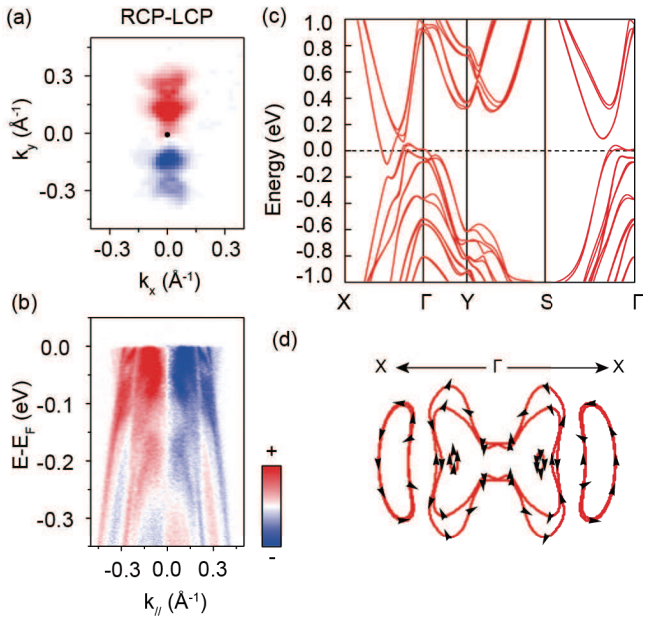}
\caption{(a) The difference of the Fermi surface map of WTe$_2$ measured under RCP and LCP light. The intensity was integrated over a window of [$E_F$ - 10~meV, $E_F$ + 10 ~meV]. (b) The corresponding intensity plot along the $\Gamma$-$X$ direction. (c) Density functional calculation of the electronic structure of WTe$_2$. (d) Density functional calculation of the spin structures at $E_F$. ARPES data were taken at 7~K with 58~eV photons.} \label{CD}
\end{figure}

Figure~\ref{CD}(a) show the Fermi surface mapping of the difference under the RCP and LCP light, namely, RCP-LCP. (the normalized (RCP-LCP)/(RCP+LCP) data can be found in the SM.) The photoemission intensity presents a strong intensity inversion between the RCP and the LCP data, which is most likely due to different OAM. This can be further confirmed by the corresponding  photoemission intensity plot along the $\Gamma$-$X$ direction  in Fig.~\ref{CD}(b). To  understand the CD, we have calculated the spin structure on the Fermi surface in Fig.~\ref{CD}(d), and the corresponding  electronic structure is shown in Fig.~\ref{CD}(c). As expected, the  spin orientation is antisymmetric about the $\Gamma$ point.  This is consistent with the CD data. However, one should notice that the calculated electronic structure only partially reproduces the experimental bands and Fermi surface. There are some disagreements in the details around $E_F$. During the calculation, we have noticed that the calculated electronic structure is extremely sensitive to the atomic coordinates. Further tuning of various parameters are still in progress to fit the experimental band structure and Fermi surface. Nevertheless, we emphasize that the strong SOC effects and spin texture is always present in the calculations.

\begin{figure}
\includegraphics[width=8.7cm]{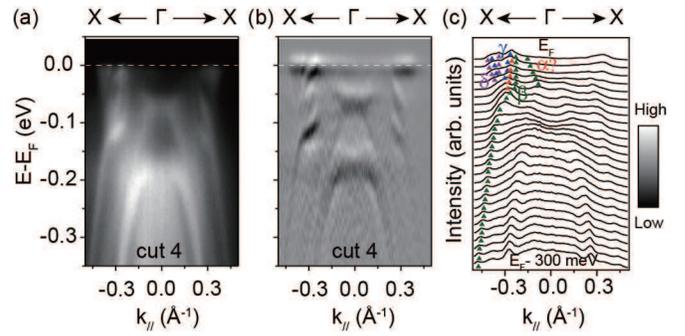}
\caption{(a)Electronic structure of another surface observed. (b) The second derivative spectrum of panel (a). (c) The corresponding MDC's of the spectrum in panel (a). Data  were taken at 9~K with 21.2~eV photons.
} \label{FS2}
\end{figure}

Intriguingly, on some of the samples, the central hole pocket around $\Gamma$ is absent [Figs.~\ref{FS2}(a) and (b)]. The only difference between the them is the states around $E_F$, while the other features of the electronic structures do not show any differences (see SM). This may be attributed to the sensitive structure-electronic structure dependence as found in our calculations. The two electron bands $\gamma$ and $\delta$ can be resolved from the corresponding MDC's in Fig.~\ref{FS2}(c). Due to the complexity of the bands around $E_F$, the hole band $\alpha$ is discernible. Despite the difference between the two resolved surfaces, this kind of surface shows the same CD, either (see SM).

Both our CD-ARPES data and density functional calculations have revealed the opposite spin and OAM structure with respect to the $\Gamma$ point. Consequently, this would suppress the backscattering channel of the quasiparticles, and thus reduces the resistivity. Naturally,  magnetic field would open up backscattering channels and increase the resistivity, and thus induce MR effect.  This is analogous to the arguments for the MR in Cd$_3$As$_2$ \cite{Cava_CdAs}. On the other hand, since the ratio of the populations of the holes and electrons is about 0.89, the electrons and holes are nearly compensate within our experimental error bar, our findings may suggest that both mechanism may conspire the anomalously large and non-saturating MR of WTe$_2$.


In conclusion, we have presented   comprehensive  electronic structure of WTe$_2$. Particularly, the strong CD indicated an exotic spin structure of this material, which is further confirmed by our band  calculations. Our results suggest that in addition to the electron-hole compensation reported earlier,  spin-orbital coupling may also play an important  role in determining the MR of WTe$_2$, which would facilitate a more comprehensive understanding of  the MR in WTe$_2$.

We thank the support from Dr. P. Dudin, Dr. T. Kim and Dr. M. Hoesch at DLS, Dr. P. Kumar Das and Dr. I. Vobornic at Elettra, and helpful discussion with Prof. Shiyan Li at Fudan University. This work is supported in part by the National Science Foundation of China and National Basic Research Program of China (973 Program) under the grant Nos. 2012CB921402, 2011CB921802, 2011CBA00112.

\end{document}